\documentclass[12pt]{article}

\usepackage{sbc-template}

\usepackage{graphicx,url}
\usepackage{listings}
\usepackage{xcolor}

\usepackage[brazil]{babel}   
\usepackage[utf8]{inputenc}

\usepackage[square,sort,comma,numbers]{natbib}

\definecolor{codegreen}{rgb}{0,0.6,0}
\definecolor{codegray}{rgb}{0.5,0.5,0.5}
\definecolor{codepurple}{rgb}{0.58,0,0.82}
\definecolor{numberbg}{rgb}{0.95,0.95,0.95}    
\definecolor{ballblue}{rgb}{0.26, 0.67, 0.8} 
\definecolor{numbercolor}{rgb}{0.4, 0.4, 0.4} 
\definecolor{graybg}{rgb}{0.95,0.95,0.95}

\title{Digital identity management system with blockchain:\\
An implementation with Ethereum and Ganache
}

\author{André Davi Lopes\inst{1}, Tais Mello\inst{1}, Wesley dos Reis Bezerra\inst{1}}

\address{Instituto Federal Catarinense -- Campus Rio do Sul
  \email{\{andredavilopes6, taismello204\}@gmail.com, wesley.bezerra@ifc.edu.br}
}

\begin{document} 

\maketitle

\begin{abstract}
This paper presents the development of a distributed digital identity system utilizing modern technologies, including FastAPI, MongoDB, gRPC, Docker, and blockchain simulation with Ganache and Ethereum. The objective is to demonstrate the benefits of distributed systems and blockchain for the security, traceability, and decentralization of digital identities. The methodology included the development of a microservices architecture with JWT authentication, data persistence in MongoDB, simulation of blockchain operations using Ganache, and containerization with Docker. The results demonstrate the feasibility of the proposed approach, with a functional web interface, complete audit logs, and blockchain simulation with Ethereum. The theoretical foundations, technical implementation, results obtained, and prospects for integration with real blockchain networks are discussed.
\end{abstract}
     

\section{Introduction}
\label{sec_instroduction}

Digital transformation has driven the need for secure and efficient solutions for digital identity management (Zyskind et al., 2015). In a scenario where personal information is exchanged in large quantities and is decentralized, ensuring data authenticity, privacy, and integrity has become a central challenge for both organizations and users. Traditional identity systems, based on centralized databases, present vulnerabilities such as single points of failure, hacking risks, and auditing difficulties.

In this context, emerging technologies such as blockchain (Nakamoto, 2008) and distributed systems (Tanenbaum \& Van Steen, 2007) have been explored to improve security, transparency, and traceability in digital identity management. Blockchain, due to its immutable and decentralized nature, enables the reliable recording of transactions and information, reducing fraud and increasing trust between the parties involved. Furthermore, the adoption of distributed architectures provides greater availability and fault tolerance, essential characteristics for critical applications.

This work presents the development of a distributed digital identity system utilizing web technologies, including the FastAPI framework (FastAPI Documentation), MongoDB (MongoDB Documentation), and blockchain simulation with Ganache and Ethereum (Ganache Documentation). The objective is to demonstrate, in practice, how the integration of these technologies can contribute to the construction of more secure and auditable solutions, discussing the main concepts, challenges, and results obtained during the system's development.

The remainder of the article is organized as follows: Section \ref{sec_literature_review} presents a literature review on the topic, providing the authors' perspective. Next, Section \ref{sec_methodology} presents the methodology employed in this work, followed by the development section (Section \ref{sec_development}), which details the development process. Equally important, Section \ref{sec_results} presents the results and discussions of the work. Finally, Section \ref{sec_conclusion} presents the conclusions of the work.

\section{Literature Review}
\label{sec_literature_review}

The theoretical foundation of this work addresses the main concepts and technologies used during the development of the digital identity system. The fundamentals of blockchain, modern web systems, distributed architectures, gRPC communication, data persistence with MongoDB, and containerization practices with Docker are presented. Understanding these topics is crucial for comprehending the design decisions, challenges encountered, and solutions implemented during the development of the proposed system.

\subsection{Blockchain}
\label{subsec_blockchain}

Blockchain is a distributed ledger technology that enables the secure and transparent storage of information in a decentralized network (Nakamoto, 2008). Based on cryptography and distributed consensus, blockchain offers unique features such as immutability, transparency, and decentralization (Zyskind et al., 2015).

The basic structure of a blockchain consists of chained blocks, where each block contains validated transactions and a hash that references the previous block, creating an immutable chain. Consensus among network nodes guarantees data integrity, while cryptography ensures the authenticity of transactions. These characteristics make blockchain ideal for applications that require high reliability and traceability, such as digital identity systems.

\subsection{Web Systems}
\label{subsec_websystems}

Modern web systems follow distributed and scalable architectures, using frameworks that facilitate the development of RESTful APIs and integration with different technologies (FastAPI Documentation). FastAPI, for example, is a modern Python framework that offers high performance, automatic typing, and automatic API documentation.

Authentication in modern web systems often utilizes JWT tokens (JSON Web Tokens), which enable identity verification without requiring server-side state storage. This approach is advantageous in distributed architectures, where multiple services can independently validate a user's authenticity.

\subsection{Distributed Systems}
\label{subsec_distributedsystems}

Distributed systems are composed of multiple autonomous components that cooperate to achieve a common goal (Tanenbaum \& Van Steen, 2007). These systems offer characteristics such as transparency, scalability, and fault tolerance, but they also introduce significant challenges related to consistency, availability, and network partitioning.

The main challenges in distributed systems include data consistency, where multiple copies of information must remain synchronized; System availability, ensuring that the service remains accessible even in the event of failures; and fault tolerance, allowing the system to continue functioning even when some components fail.

\subsection{gRPC API}
\label{subsec_grpc}

gRPC (Google Remote Procedure Call) is an inter-service communication framework developed by Google that uses the HTTP/2 protocol and the Protocol Buffers serialization format (gRPC Documentation). This technology offers high performance, strong typing, and bidirectional streaming support, making it especially suitable for communication between microservices in distributed systems.

gRPC allows the definition of service interfaces through .proto files, which are then used to generate client and server code in multiple programming languages. This approach ensures consistency between different implementations and facilitates integration between heterogeneous services.

\subsection{MongoDB}
\label{subsec_mongdb}

MongoDB is a document-oriented NoSQL database that offers high scalability and flexibility (MongoDB Documentation). Its distributed architecture allows for automatic replication and data distribution.

MongoDB utilizes a JSON document-based data model, providing flexibility to store identity information in various structures. Its automatic replication and data distribution capabilities make it suitable for applications that require high availability and eventual consistency, important characteristics in distributed digital identity systems.

\subsection{Docker and Containerization}
\label{subsec_docker}

Docker is a containerization platform that allows you to package applications and their dependencies in isolated containers (Docker Documentation). This technology has revolutionized software development and deployment, offering portability, consistency, and resource efficiency.

Docker containers provide an isolated and reproducible environment, ensuring that the application works the same way across different environments (development, testing, production). This feature is particularly valuable in distributed systems, where multiple services must be orchestrated and managed efficiently.

Docker Compose is a tool that allows you to define and run multi-container applications through YAML files. This approach significantly simplifies the configuration and orchestration of complex systems, allowing developers to define all the necessary services in a single configuration file.

Containerization offers several advantages for digital identity systems: service isolation, ease of deployment, horizontal scalability, and simplified dependency management. These features make Docker a crucial technology for developing modern, distributed systems.

\section{Methodology}
\label{sec_methodology}

The methodology adopted in this work is based on the development of a distributed digital identity system using modern technologies and agile development practices. The project was structured in iterative phases, allowing for incremental development and continuous validation of the implemented functionalities.

\subsection{Chosen Technologies}
\label{subsec_techonologies}

The selection of technologies was based on criteria of modernity, suitability for the problem, and ease of implementation. FastAPI was chosen as the main \textit{web} framework due to its high \textit{performance}, automatic typing, and automatic API documentation (FastAPI Documentation). MongoDB was selected as the database due to its flexibility in storing JSON documents and replication capabilities (MongoDB Documentation).

For the blockchain simulation, we chose to use Ganache with Ethereum, which enables the simulation of a local blockchain network in a simplified manner, demonstrating the concepts practically and educationally (Ganache Documentation). Communication between services was implemented using gRPC (gRPC Documentation), which offers high performance and strong typing for communication between microservices.

\subsection{Proposal Architecture}
\label{sec_architecture}

\begin{figure}[ht]
\centering
\includegraphics[width=.7\textwidth]{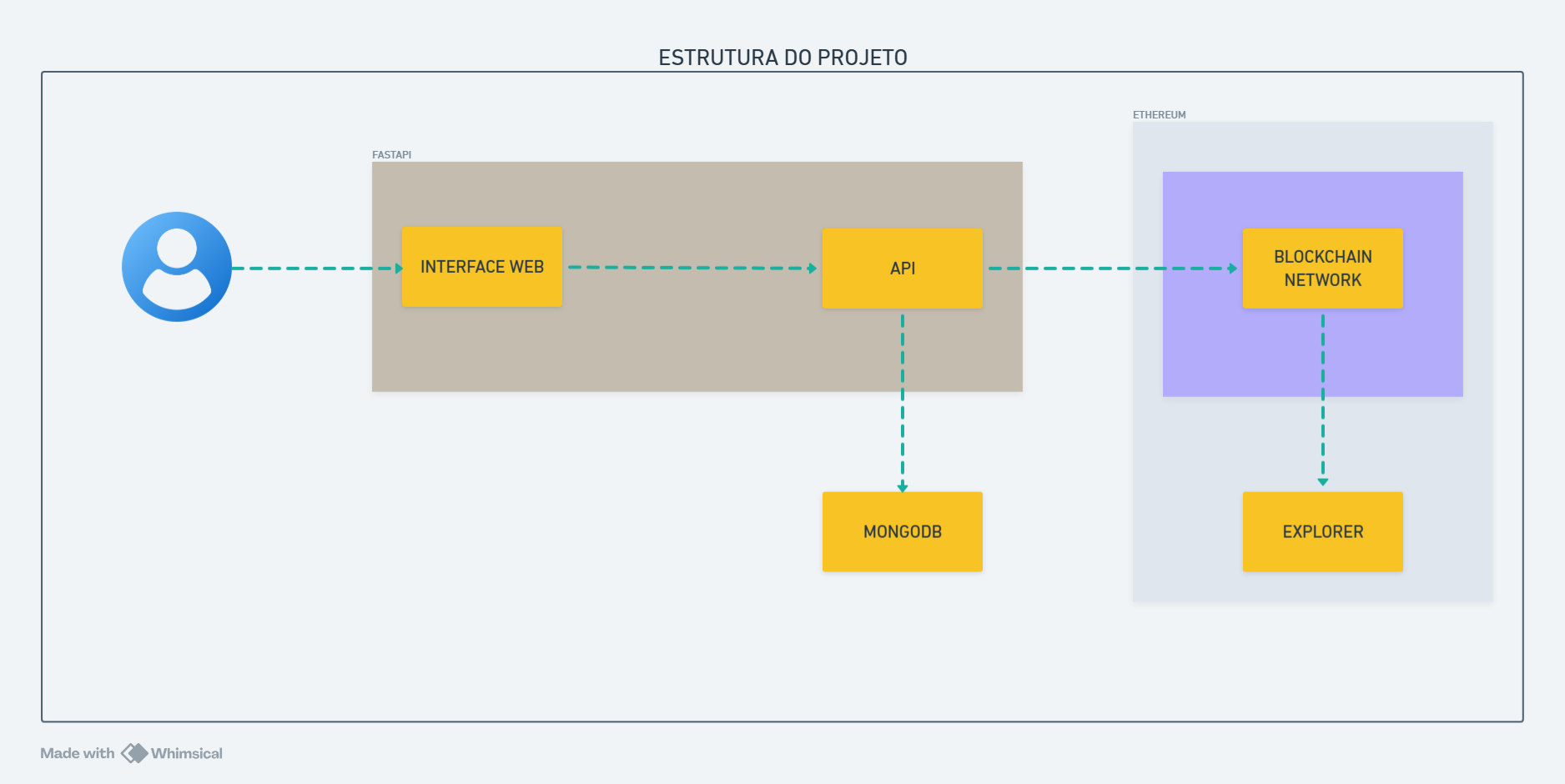}
\caption{System Architecture}
\label{fig:project_structure}
\end{figure}

Figure \ref{fig:project_structure} presents the overall system architecture, illustrating the interaction between the various components and the communication established between them.

The system was designed following a microservices architecture, where each component has well-defined responsibilities. The main application (FastAPI) acts as an inbound gateway, managing authentication, authorization, and request routing. The gRPC service simulates blockchain operations with Ganache, while MongoDB stores user data and audit logs.

Docker containerization was used to ensure portability and facilitate system deployment. Docker Compose was used to orchestrate the multiple services, simplifying the configuration and execution of the development environment.

\subsection{Main Design Decisions}
\label{subsec_design_decision}

One of the key decisions was to use Ganache to simulate the Ethereum blockchain instead of a real network, allowing us to focus on fundamental concepts without the complexity of setting up a production blockchain network. This approach facilitated the development and demonstration of concepts while maintaining the possibility of future integration with a real Ethereum network.

JWT-based authentication was chosen for its suitability in distributed architectures, allowing multiple services to validate user authenticity independently. Data persistence was implemented in MongoDB, leveraging its flexibility to store documents with different structures.

\section{Development}
\label{sec_development}

The system was developed using a modular approach, with each component being implemented and tested independently before integration. The implementation spans the web interface to the simulation of blockchain operations, encompassing authentication and data persistence.

\subsection{System Structure}
\label{subsec_system_structure}

The system was structured into three main components: the web frontend, the backend API, and the blockchain simulation. The frontend was developed using HTML, CSS, and JavaScript, providing a modern and responsive interface for user interaction. The backend was implemented in Python using the FastAPI framework (FastAPI Documentation), offering high performance and automatic API documentation.

The blockchain simulation was implemented using Ganache with Ethereum, which simulates the operations of a real blockchain network, including transaction recording, hash generation, and maintenance of a distributed ledger. This approach allows for a practical and educational demonstration of the fundamental concepts of blockchain (Ganache Documentation).

\subsection{Authentication and Authorization Flow}
\label{subsec_auth_flow}

Authentication was implemented using JSON Web Tokens (JWT), which enable identity verification without requiring server-side state storage. The authentication flow includes:

\begin{enumerate}
\item User registration with data validation (CPF, email)
\item \textit{Login} with credential verification
\item Generation of JWT token with expiration time
\item Validation of the token in subsequent requests
\item \textit{Logout} with token invalidation
\end{enumerate}

Authorization was implemented through \textit{middleware} that verifies the presence and validity of the JWT token in all protected routes. This approach ensures that only authenticated users can access sensitive system features.

The JWT authentication implementation was performed as shown in the example below:

\begin{lstlisting}[language=Python, caption={JWT Implementation}]

from fastapi import Depends, HTTPException, status
from fastapi.security import HTTPBearer
import jwt

security = HTTPBearer()

def create_access_token(data: dict, expires_delta: timedelta):
    to_encode = data.copy()
    expire = datetime.utcnow() + expires_delta
    to_encode.update({"exp": expire})
    encoded_jwt = jwt.encode(to_encode, SECRET_KEY, algorithm="HS256")
    return encoded_jwt

def get_current_user(token: str = Depends(security)):
    credentials_exception = HTTPException(
        status_code=status.HTTP_401_UNAUTHORIZED,
        detail="Could not validate credentials"
    )
    try:
        payload = jwt.decode(token, SECRET_KEY, algorithms=["HS256"])
        email: str = payload.get("sub")
        if email is None:
            raise credentials_exception
    except JWTError:
        raise credentials_exception
    return email
\end{lstlisting}

The most important parts of the JWT implementation include: the \texttt{create\_access\_token} function, which generates \textit{tokens} with expiration times; the \texttt{get\_current\_user} function, which validates \textit{tokens} and extracts user information; and exception handling, which ensures security in the event of invalid \textit{tokens}.

\subsection{Data Persistence in MongoDB}

Data persistence was implemented using MongoDB (MongoDB Documentation), taking advantage of its flexibility to store JSON documents with different structures. The system uses two main collections:

\begin{itemize}
\item \textbf{users}: Stores user information (name, CPF, email, password \textit{hash})
\item \textbf{audit\_logs}: Records all actions performed in the system for auditing purposes
\end{itemize}

\begin{figure}[ht]
\centering
\includegraphics[width=.8\textwidth]{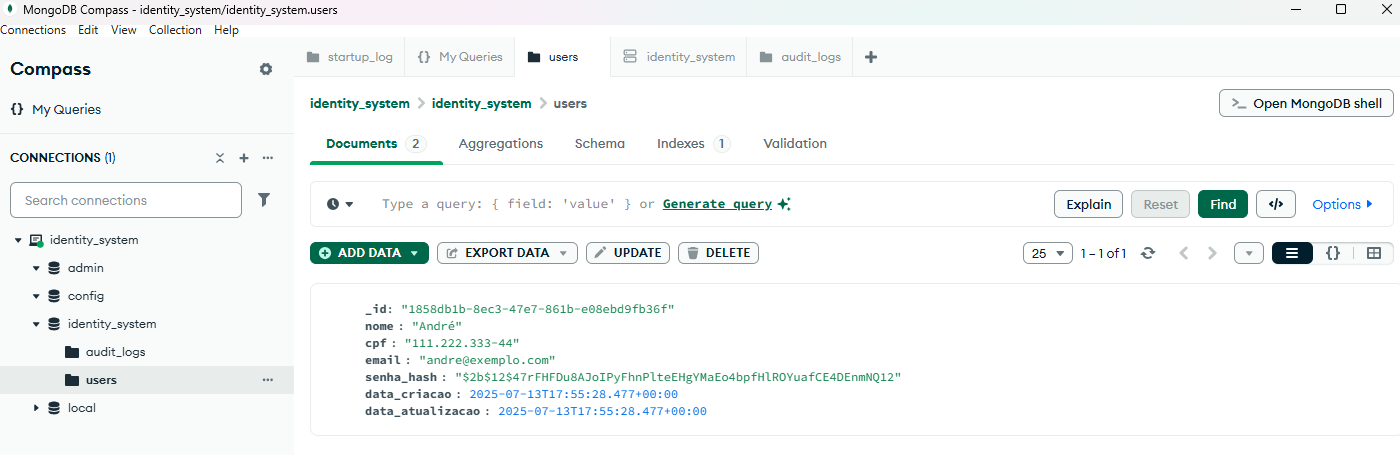}
\caption{Data Persistence in MongoDB}
\label{fig:mongo_db}
\end{figure}

Figure \ref{fig:mongo_db} demonstrates how data is persisted in MongoDB, showing the collection structure and how information is stored in an organized and accessible manner.

The connection to MongoDB was implemented using the PyMongo library, with error handling and automatic reconnection in case of failures. Data validation was implemented using Pydantic, ensuring the integrity of the input data.

Example implementation of the connection to MongoDB:

\begin{lstlisting}[language=Python, caption={MongoDB Connection}]
# Conexao com MongoDB
from pymongo import MongoClient
from pymongo.errors import ConnectionFailure
        
    async def connect(self):
        try:
            self.client = MongoClient("mongodb://mongodb:27017/")
            self.db = self.client.identity_system
            print("Conectado ao MongoDB")
        except ConnectionFailure as e:
            print(f"Erro ao conectar ao MongoDB: {e}")
\end{lstlisting}

Key points of the MongoDB implementation include the \texttt{connect} method (lines 5-11), which establishes the connection with error handling.

\subsection{Blockchain Operations}

\textit{blockchain} operations were implemented using Ganache with Ethereum, which simulates the fundamental operations of a real \textit{blockchain} network. The service includes:

\begin{itemize}
\item \textbf{Transaction Log}: Simulates sending data to the Ethereum blockchain
\item \textbf{Hashes Generation}: Implements hashing algorithms to ensure integrity
\item \textbf{Distributed Ledger}: Keeps a record of transactions made in Ganache
\item \textbf{Transaction Validation}: Verifies data authenticity using smart contracts
\end{itemize}

\begin{figure}[ht]
\centering
\includegraphics[width=.9\textwidth]{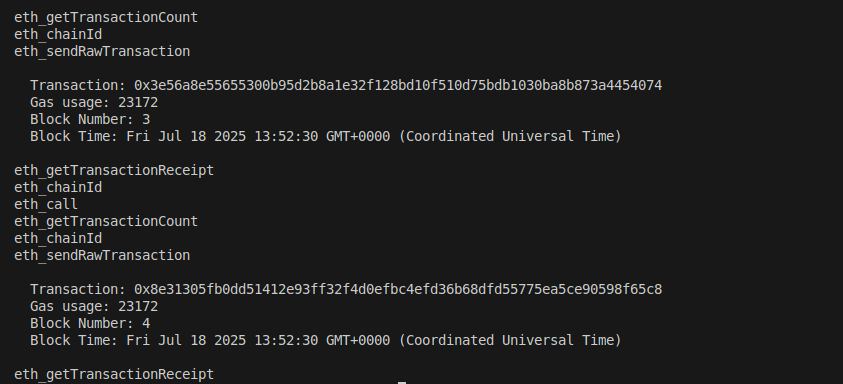}
\caption{ Logs returned by the Blockchain API with Ganache}
\label{fig:interface_blockchain}
\end{figure}

Figure \ref{fig:interface_blockchain} shows the \textit{blockchain} simulation interface with Ganache, where you can see how transactions are recorded and how the system simulates the operation of a local Ethereum network.

It is essential to note that this implementation utilizes Ganache for educational and development purposes. In a production environment, it is necessary to integrate with a real Ethereum network or a test network, such as Rinkeby or Goerli.

Blockchain implementation with Ganache:

\begin{lstlisting}[language=Python, caption={Blockchain implementation using Ganache and Ethereum}]
    def registrar_usuario_blockchain(self, user_data: dict) -> dict:
        try:
            CONTA_ADMIN = Web3.to_checksum_address(self.w3.eth.accounts[0])
            
            tx = self.contract.functions.registrarUsuario(
                user_data["user_id"], 
                user_data["email"], 
                user_data["cpf"]
            ).build_transaction({
                'from': CONTA_ADMIN,
                'nonce': self.w3.eth.get_transaction_count(CONTA_ADMIN),
                'gas': 3000000,
                'gasPrice': self.w3.to_wei('20', 'gwei')
            })
            
            signed_tx = self.w3.eth.account.sign_transaction(tx, private_key=PRIVATE_KEY)
            tx_hash = self.w3.eth.send_raw_transaction(signed_tx.raw_transaction)
            receipt = self.w3.eth.wait_for_transaction_receipt(tx_hash)
            
            return {
                "transaction_hash": receipt.transactionHash.hex(),
                "block_number": receipt.blockNumber,
                "gas_used": receipt.gasUsed
            }
        except Exception as e:
            return {"error": str(e)}
\end{lstlisting}

In Algorithm 3, the \texttt{register\_user\_blockchain} method, responsible for executing real transactions on the Ethereum network, encompasses transaction construction, signing, sending, and block confirmation.

The implementation utilizes real smart contracts on the Ethereum network, as evidenced by the \texttt{logs}, which record operations such as \texttt{eth\_sendRawTransaction} and \texttt{eth\_getTransactionReceipt}. Transactions are confirmed in specific blocks, and actual \texttt{gas} consumption is reported (23172 units per transaction).

\subsection{Containerization with Docker}

Containerization was implemented using Docker, allowing the system to run in any environment that supports \textbf{containers}. Docker Compose was used to orchestrate the multiple services:

\begin{itemize}
\item \textbf{fastapi-app}: \textit{Container} for the main application
\item \textbf{mongodb}: \textit{Container} for the database
\item \textbf{ganache-simulator}: \textit{Container} for the \textit{blockchain} simulation with Ganache
\end{itemize}

The Docker Compose configuration includes defining networks, persistent volumes, and environment variables, facilitating \textit{deployment} and system maintenance. Communication between \textit{containers} is carried out through a dedicated Docker network, ensuring isolation and security.

\subsection{Web Interface and User Experience}

\begin{figure}[ht]
\centering
\includegraphics[width=.7\textwidth]{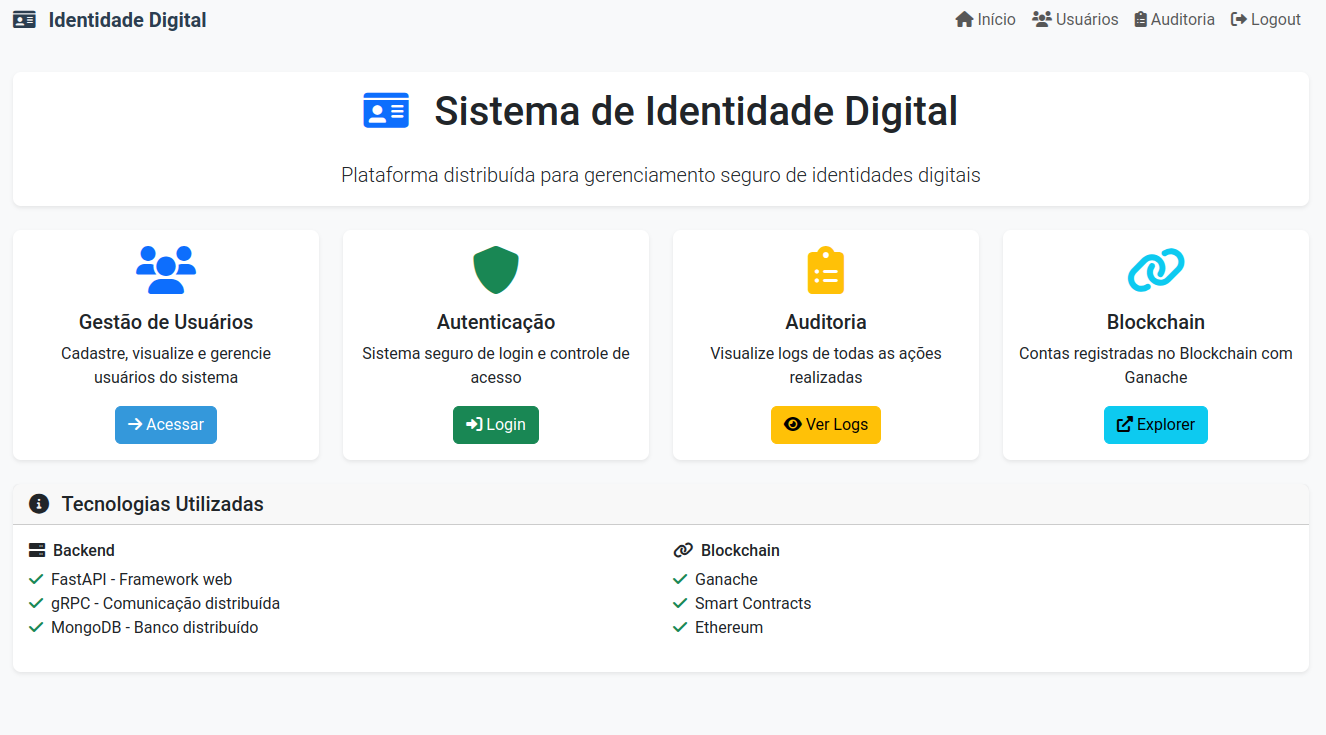}
\caption{Main System Interface}
\label{fig:interface_principal}
\end{figure}

The \textit{web} interface was developed with a focus on usability and accessibility, including pages for:

\begin{itemize}
\item \textbf{\textit{Login} and Registration}: Forms with real-time validation
\item \textbf{User Management}: Listing, editing, and deleting users
\item \textbf{\textit{Audit Logs}}: View all actions performed in the system
\item \textbf{Responsive Navigation}: \textit{Menu} adaptable to different Devices
\end{itemize}

The interface uses Bootstrap for responsive styling and JavaScript for dynamic interactions, providing a modern and intuitive user experience.

Figure \ref{fig:interface_principal} shows the system's main screen, showing the available features and navigation between the different sections.

\section{Results and Discussions}
\label{sec_results}

The implementation of the distributed digital identity system allowed for the practical validation of the concepts of security, traceability, and decentralization discussed in the theoretical framework. The simulation environment, based on Docker Compose and Ganache, enabled the integrated execution of the backend, database, and Ethereum blockchain simulation services, facilitating development and testing.

\subsection{System Operation Demonstration}
\label{subsec_system_demonstration}

The system ran locally, enabling user registration, authentication, and management through a responsive web interface. All sensitive operations, including user creation, editing, and deletion, were recorded in audit logs, which could be accessed directly from the interface. The blockchain simulation with Ganache recorded critical transactions, illustrating how Ethereum technology can be used to ensure data integrity and traceability.

JWT authentication worked as expected, protecting sensitive routes and ensuring that only authenticated users had access to restricted features. The use of MongoDB enabled efficient data persistence, with fast and flexible queries.

\subsection{Observed Benefits}
\label{subsec_benefits}

The main benefits observed during system development and testing include:

\begin{itemize}
\item \textbf{Security}: JWT authentication and audit logging increased protection against unauthorized access and facilitated the detection of suspicious actions.
\item \textbf{Traceability}: All relevant actions were logged, allowing for complete auditing of system operations.
\item \textbf{Ease of Deployment}: The use of Docker and Docker Compose simplified the configuration and execution of the environment, making the system easily replicable.
\item \textbf{Flexibility}: The modular architecture allowed for the replacement or enhancement of components (e.g., migrating Ganache to a real Ethereum network in the future). \end{itemize}

These benefits demonstrate the robustness and adaptability of the developed solution, reinforcing its potential to evolve and be integrated into more complex or demanding environments.

\subsection{Limitations and Future Perspectives}

Despite the positive results, some limitations were identified:

\begin{itemize}
\item \textbf{Blockchain Simulation}: The use of Ganache limits the guarantee of immutability and distributed consensus to a local environment. For production environments, integration with real or test Ethereum networks is recommended (Ethereum Documentation).
\item \textbf{Real Scalability}: The tests were conducted in a local environment; performance was not evaluated in high-demand scenarios or real distributed environments. 
\item \textbf{Advanced Features}: Features such as multi-factor authentication, integration with external identity providers, and the use of real Ethereum smart contracts can be explored in future work.
\end{itemize}

Overall, the developed system met its proposed objectives, demonstrating the potential of combining distributed systems, modern authentication, and the Ethereum blockchain (even when simulated with Ganache) for the secure management of digital identities.

\section{Conclusion}
\label{sec_conclusion}
The development of a distributed digital identity system using Ganache to simulate the Ethereum blockchain enabled us to demonstrate in practice the benefits and challenges of utilizing modern technologies, including FastAPI, MongoDB, gRPC, and blockchain. The proposed architecture proved efficient in ensuring security, traceability, and flexibility in digital identity management, meeting the requirements for authentication, auditing, and data persistence. The use of JWT-based authentication and detailed audit logs contributed to increased system security and transparency. Simulating blockchain operations with Ganache enabled an understanding of the concepts of immutability and decentralization using Ethereum technology, even without the complexity of a real network.

Despite the limitations inherent in simulation, the developed system serves as a solid foundation for future developments, including integration with real Ethereum networks, the implementation of smart contracts, and the adoption of advanced authentication mechanisms.

Future work suggests testing in real-world distributed environments, conducting large-scale performance evaluations, and exploring integrations with external identity providers.

In short, this work highlights the potential of combining distributed systems, modern authentication, and the Ethereum blockchain to build secure and auditable digital identity solutions, thereby contributing to the advancement of the field and paving the way for new research and applications.



\begin{thebibliography}{99}
\bibitem{ganache} Ganache Documentation. Available em: \url{https://trufflesuite.com/ganache/}
\bibitem{ethereum} Ethereum Documentation. Available em: \url{https://ethereum.org/en/developers/docs/}
\bibitem{fastapi} FastAPI Documentation. Available em: \url{https://fastapi.tiangolo.com/}
\bibitem{mongodb} MongoDB Documentation. Available em: \url{https://www.mongodb.com/docs/}
\bibitem{blockchain-id} Zyskind, G., Nathan, O., \& Pentland, A. (2015). Decentralizing privacy: Using blockchain to protect personal data. In 2015 IEEE Security and Privacy Workshops.
\bibitem{nakamoto} Nakamoto, S. (2008). Bitcoin: A peer-to-peer electronic cash system.
\bibitem{tanenbaum} Tanenbaum, A. S., \& Van Steen, M. (2007). Distributed Systems: Principles and Paradigms. Prentice Hall.
\bibitem{allen} Allen, C. (2016). The path to self-sovereign identity. Life with Alacrity.
\bibitem{grpc} gRPC Documentation. Available em: \url{https://grpc.io/docs/}
\bibitem{docker} Docker Documentation. Available em: \url{https://docs.docker.com/}
\end{thebibliography}
\end{document}